\documentclass[journal]{IEEEtran}
\usepackage{amsfonts}
\usepackage{amssymb}
\usepackage{stfloats}
\usepackage{cite}
\usepackage{graphicx}
\usepackage{psfrag}
\usepackage{tikz}
\usetikzlibrary{math}
\usepackage{subfigure}
\usepackage{amsmath}
\usepackage{array}
\usepackage{epstopdf}
\usepackage{authblk}
\usepackage{graphicx} 
\usepackage{amsthm} 
\usepackage{lipsum}
\usepackage{verbatim} 
\usepackage{color}
\usepackage{caption}
\usepackage{amsmath,amsfonts,amssymb,mathrsfs}
\usepackage{bm}
\usepackage{ dsfont }
\usepackage{ upgreek }
\usepackage{multirow,enumerate}

\usepackage{pifont}
\usepackage{graphicx}

\newtheorem{Remark}{Remark}

\usepackage{algorithm}
\usepackage{algorithmicx}
\usepackage{algpseudocode}

\begin{document}
	\title{\huge On DoF of Active RIS-Assisted MIMO Interference Channel with Arbitrary Antenna Configurations: When Will RIS Help?
	}
	\allowdisplaybreaks[4]
	\author{Shuo Zheng, Bojie Lv, Tong Zhang,  Yinfei Xu,   Gaojie Chen, \textit{Senior Member, IEEE}, \\  Rui Wang, \textit{Member, IEEE},  and P. C. Ching, \textit{Life Fellow, IEEE,}

%		\thanks{Manuscript received December 1, 2022; revised April 17, 2023; accepted July 6, 2023. This work was supported by the National Natural Science Foundation of China under Grant 62171213, Fundamental Research Funds for the Central Universities under Grant XXXXXX, and  Zhi Shan Young Scholar Program of Southeast University.  The associate editor coordinating the review of this paper and approving it for publication was Dr. Yuan Wu \emph{(Corresponding author:  Rui Wang, Tong Zhang).}}
	
		\thanks{
			S. Zheng, B. Lv, and R. Wang are with the Department of Electrical and Electronic Engineering, Southern University of Science and Technology, Shenzhen 518055, China (\{zhengs2021, lyubj\}@mail.sustech.edu.cn, wang.r@sustech.edu.cn).
			
			T. Zhang is with Department of Electronic Engineering, College of Information Science and Technology, Jinan University, Guangzhou 510632, China, and also with the Department of Electrical and Electronic Engineering, Southern University of Science and Technology, Shenzhen 518055, China (e-mail: zhangt77@jnu.edu.cn).
			
			Y. Xu is with School of Information Science and Engineering, Southeast
			University, Nanjing, 210096, China (yinfeixu@seu.edu.cn).
			
			%S. Wang is with the Guangdong-Hong Kong-Macao Joint Laboratory of
			%Human-Machine Intelligence-Synergy Systems, Shenzhen Institute of Advanced
			%Technology, Chinese Academy of Sciences, Shenzhen, 518055, China
			%(e-mail: s.wang@siat.ac.cn).
			
			G. Chen is with 5GIC \& 6GIC, Institute for Communication Systems, University of Surrey, Guildford GU2 7XH, UK (gaojie.chen@surrey.ac.uk).
			
			P. C. Ching is with the Department of Electronic Engineering, The Chinese University of Hong Kong, Hong Kong (pcching@ee.cuhk.edu.hk).
	}}
	
	\maketitle

	\begin{abstract}
		An active reconfigurable intelligent surface (RIS) has been shown to be able to enhance the sum-of-degrees-of-freedom (DoF) of a two-user multiple-input multiple-output (MIMO) interference channel (IC) with equal number of  antennas at each transmitter and receiver. However, for any number of receive and transmit antennas, when and how an active RIS can help to improve the sum-DoF are still unclear. This paper studies the sum-DoF of an active RIS-assisted two-user MIMO IC with arbitrary antenna configurations. In particular, RIS beamforming, transmit zero-forcing, and interference decoding are integrated together to combat the interference problem. In order to maximize the achievable sum-DoF, an integer optimization problem is formulated to optimize the number of eliminating interference links by RIS beamforming.  As a result, the derived achievable sum-DoF can be higher than the sum-DoF of two-user MIMO IC, leading to a RIS gain. Furthermore, a sufficient condition of the RIS gain is given as the relationship between the number of RIS elements and the antenna configuration. 

	\end{abstract}
	
	\begin{IEEEkeywords}
		DoF, RIS, MIMO interference channel
	\end{IEEEkeywords}
	\section{Introduction}
	
	Reconfigurable intelligent surface (RIS) emerges as a revolutionary tool to ameliorate the communication environment \cite{ycs,Han,Renzo}. Specifically, the active RIS can adjust both the amplitude and phase of each RIS element to reflect the incident signal  \cite{zhang2022multi}.  Unfortunately, the exact channel capacity of RIS-assisted multi-user communications is, however, hard to obtain.  Therefore, the degrees-of-freedom (DoF), i.e., the maximal number of interference-free channels, which is a first-order approximation of the channel capacity in a high signal-to-noise ratio (SNR) regime, has become an alternative tractable solution.
	
	The DoF for RIS-assisted multi-user communications has been investigated in \cite{ywisit,ywtit,non, Mohamed, activepassive,  co}. The authors in \cite{ywisit} and \cite{ywtit} showed if the data stream is available at a passive RIS and can be modulated through the adjustable phases at the passive RIS, the sum-DoF can be significantly elevated in the point-to-point multiple-input multiple-output (MIMO) channel. In \cite{non}, the DoF gain was characterized by allowing a passive RIS to transmit phase-modulated symbols in non-coherent communications without channel state information at the transmitter (CSIT). In \cite{Mohamed}, an active RIS-assisted MIMO wiretap channel was investigated, which showed that secure DoF can be elevated via leakage signal cancellation by RIS. For the RIS-assisted $K$-user interference channel (IC), the role of RIS assistance was studied in \cite{activepassive} and \cite{co}. With a single antenna at each transmitter and receiver, the authors of \cite{activepassive} derived lower and upper bounds for the sum-DoF of $K$-user IC with an active RIS. Recently, with an equal number of antennas at each transmitter and receiver, the authors of \cite{co} showed that the sum-DoF of $K$-user IC can be largely elevated, compared with the case without RIS. However, all existing works on applying RIS to MIMO communications are restricted by the specific requirement of the transmit and receive antenna configurations.

		\begin{figure} 
				\centering
				\includegraphics[width=.75\linewidth]{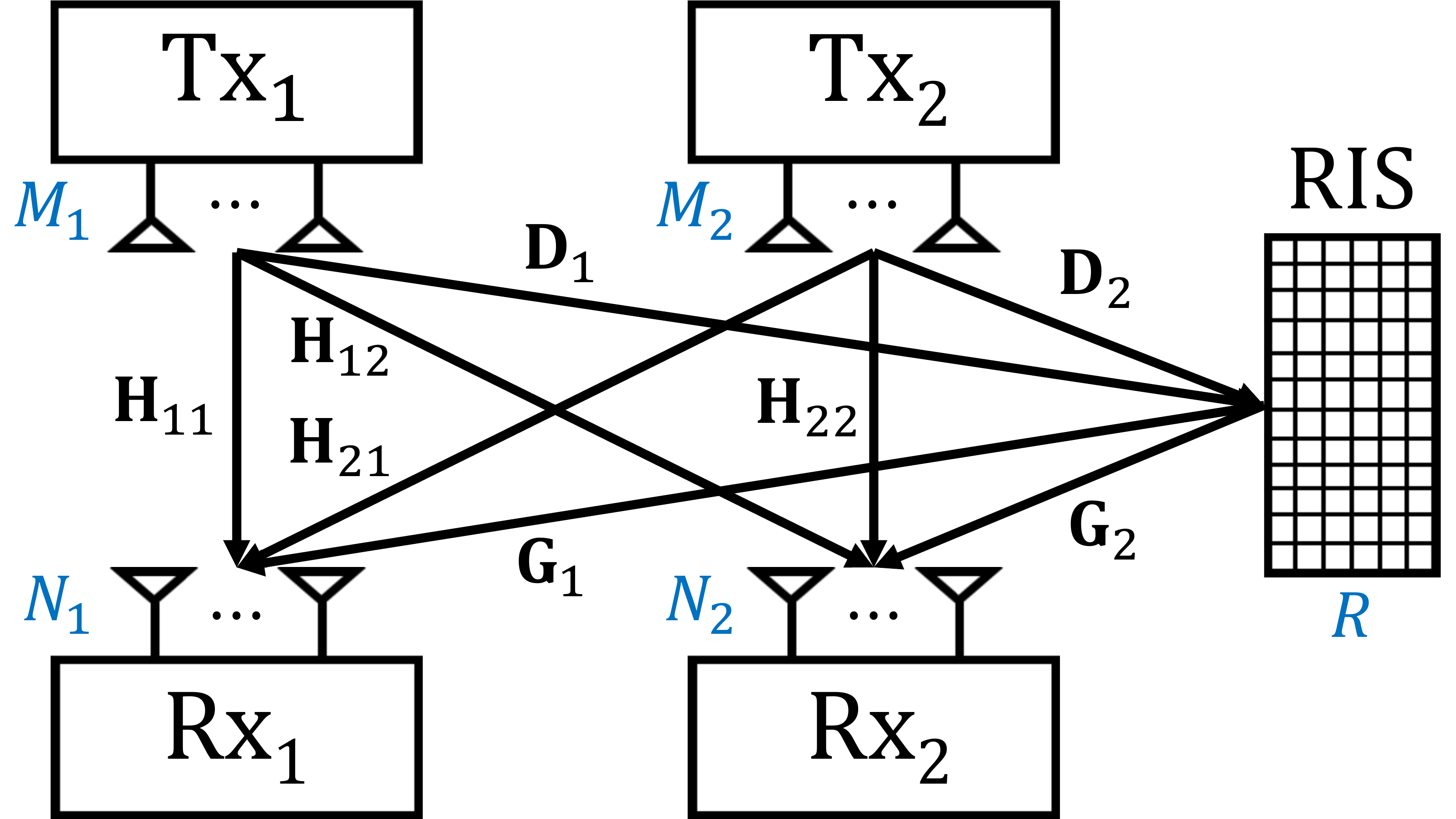}
				\caption{Illustration of the RIS-assisted two-user MIMO IC.} \label{Scenario}
				\vspace{-0.5cm}
			\end{figure}

 In this paper, in order to overcome the above limitation, we investigate the achievable sum-DoF of an active RIS-assisted two-user MIMO IC with arbitrary antenna configurations (i.e., the number of antennas of each transmitter or receiver can be arbitrary). Unlike previous works which only considered RIS beamforming \cite{ywisit,ywtit,non, Mohamed, activepassive,  co}, we further apply  transmit zero-forcing  and receive interference decoding techniques to tackle the challenges of asymmetric antenna configurations. Compared to \cite{jafar}, an active RIS is exploited to elevate the sum-DoF. The achievable sum-DoF maximization problem is formulated to obtain the optimal number of eliminating interference links by RIS beamforming. Subsequently, an optimized achievable sum-DoF together with a sufficient condition for obtaining the RIS gain are also derived. It is also shown that the RIS gain actually depends on the relationship between the number of RIS elements and the antenna configuration. Furthermore, the RIS gain in closed-form is derived in terms of symmetric antenna configurations. 

	\section{System Model and Definition}
%	\begin{figure}
%		\centering
%		\includegraphics[width=.7\linewidth]{imagefile}
%	\end{figure}
	As illustrated in Fig. \ref{Scenario}, we consider an active RIS-assisted two-user MIMO IC, where two transmitters with $M_1$ and $M_2$ antennas are denoted by $\textrm{Tx}_1$ and $\textrm{Tx}_2$ and two receivers with $N_1$ and $N_2$ antennas are denoted by $\textrm{Rx}_1$ and $\textrm{Rx}_2$, respectively. It is worth mentioning that $M_1$, $M_2$, $N_1$ and $N_2$ can be arbitrary feasible value, which is defined as arbitrary antenna configurations. Without loss of generality, we assume $\max\{M_1,N_1\}\geq  \max\{M_2,N_2\}$.
	The communication between transmitters and receivers is assisted by an active RIS equipped with $R$ reflecting elements, which can adjust the amplitude and phase shift in each element to the incident signal. 
	Transmitted signals of $\textrm{Tx}_1$ and $\textrm{Tx}_2$ are denoted by $\mathbf{P}_1\mathbf{x}_1 \in \mathbb{C}^{M_1 \times 1}$ and $\mathbf{P}_2\mathbf{x}_2 \in \mathbb{C}^{M_2 \times 1}$, respectively, where  $\mathbf{x}_1$ and $\mathbf{x}_2$ contain the messages for $\textrm{Rx}_1$ and $\textrm{Rx}_2$, respectively, and $\mathbf{P}_1$ and $\mathbf{P}_2$ are transmit beamforming matrices. The received signals at $\textrm{Rx}_1$ and $\textrm{Rx}_2$ are expressed as follows
%	\footnote{The noise term in the received signal is ignored therein and follow-up analysis, due to infinite SNR in the definition of DoF.}
	\begin{align*}
		\mathbf{y}_1 = (\mathbf{H}_{11}+\mathbf{G}_{1}\mathbf{\Psi}\mathbf{D}_{1})\mathbf{P}_1\mathbf{x}_1 + (\mathbf{H}_{21}+\mathbf{G}_{1}\mathbf{\Psi}\mathbf{D}_{2})\mathbf{P}_2\mathbf{x}_2+\mathbf{z}_1,\\
		\mathbf{y}_2 = (\mathbf{H}_{22}+\mathbf{G}_{2}\mathbf{\Psi}\mathbf{D}_{2})\mathbf{P}_2\mathbf{x}_2 + (\mathbf{H}_{12}+\mathbf{G}_{2}\mathbf{\Psi}\mathbf{D}_{1})\mathbf{P}_1\mathbf{x}_1+\mathbf{z}_2,
	\end{align*}%
	respectively, where $\mathbf{H}_{ij}\in \mathbb{C}^{N_j\times M_i},i,j=1,2$ denotes the direct channel matrix between transmitter $\textrm{Tx}_i$ and receiver $\textrm{Rx}_j$, $\mathbf{D}_{i}\in \mathbb{C}^{R\times M_i}$ denotes the channel matrix between transmitter $\textrm{Tx}_i$ and RIS, $\mathbf{G}_{i}\in \mathbb{C}^{ N_i \times R}$ denotes the channel matrix between receiver $\textrm{Rx}_i$ and RIS, $\mathbf{\Psi}=\text{diag}\{\psi_1,...,\psi_R\} \in \mathbb{C}^{R\times R}$ denotes the active RIS reflection matrix, and $\mathbf{z}_i,i=1,2$ denotes the additive white Gaussian noise (AWGN) at the receiver $\textrm{Rx}_i$. Without loss of generality, the channel matrices are assumed to be full-rank and known at both transmitters, both receivers, and the RIS.  
	
 In the IC, each $\textrm{Tx}_i, i=1,2$ has independent message $W_i,i=1,2$ desired by $\textrm{Rx}_i,i=1,2$, respectively. The rate of each message is given by $R_i(P)=\frac{\log \vert W_i(P)\vert}{n},i=1,2$, where $P$ is the power, $\vert W_i(P)\vert$ denotes the cardinality of the message set, and $n$ represents the number of channel uses. A rate tuple $(R_1(P),R_2(P))$ is said to be achievable if the decoding error probability of each message can be made arbitrarily small as channel uses tend to infinity. The capacity region $\mathcal{C}(P)$ is the closure of all achievable rates. The sum-capacity $\mathcal{C}_{\Sigma}(P)=\sup_{(R_1(P),R_2(P))\in \mathcal{C}(P)} R_1(P)+R_2(P)$ is the supremum of all achievable sum-rates. Therefore, sum-DoF is defined as $d_1+d_2=\lim\sup_{P\rightarrow \infty}\frac{\mathcal{C}_{\Sigma}(P)}{\log P}$. 

	\begin{table*}[ht]
		\centering
		\renewcommand\arraystretch{2.5}
		\caption{The Proposed Achievable Sum-DoF (Sum-DoF Lower Bound)}
		\label{DoFtable}
		\begin{tabular}{|p{0mm}c|c|c|c|}
			\hline
			&\multicolumn{2}{c}{\multirow{2}{*}{$M_1\geq \{M_2,N_1,N_2\}$ and $M_2\geq N_1$}}\vline& $R\ge  M_1M_2-M_2^2$&$\text{Sum-DoF}\geq \min\{\lfloor \frac{R+M_1^2+M_2^2}{M_1+M_2} \rfloor,M_2+N_1,N_1+N_2\}$\\
			\cline{4-5}
			 &\multicolumn{2}{c}{\hspace{4.4cm}{\footnotesize \text{Case-1}}}\vline&$R< M_1M_2-M_2^2$& $\text{Sum-DoF}\geq \min\{M_2+\lfloor\frac{R}{M_2}\rfloor,M_2+N_1,N_1+N_2\}$\\
			\hline
			&\multirow{4}{*}{\!\!\!$M_1 \geq N_2$ and $N_1 \ge M_2$}& \multirow{2}{*}{$M_1\geq N_1$}&$R\ge M_1N_1-N_1^2$ & $\text{Sum-DoF}\geq \min\{\lfloor \frac{R+M_1^2+N_1^2}{M_1+N_1} \rfloor,M_2+N_1,N_1+N_2\}$ \\
			\cline{4-5}
			&& \hspace{1.30cm}{\footnotesize  \text{Case-2-1}}&$R< M_1N_1-N_1^2$ & $\text{Sum-DoF}\geq \min\{ N_1+\lfloor\frac{R}{N_1} \rfloor,M_2+N_1,N_1+N_2\}$\\
			\cline{3-5}
			&& \multirow{2}{*}{$N_1>M_1$}& $R\ge M_1N_1-M_1^2$ & $\text{Sum-DoF}\geq \min\{\lfloor \frac{R+M_1^2+N_1^2}{M_1+N_1} \rfloor,M_1+N_2,M_1+M_2\}$\\
			\cline{4-5}
			&& \hspace{1.30cm}{\footnotesize \text{Case-2-2}}&$R< M_1N_1-M_1^2$ & $\text{Sum-DoF}\geq \min\{ M_1+\lfloor\frac{R}{M_1} \rfloor,M_1+N_2,M_1+M_2\}$\\
			\hline
			 &\multicolumn{2}{c}{\multirow{2}{*}{$N_1 \ge \{M_1,M_2,N_2\}$ and $N_2 \ge M_1$}}\vline& $R\ge N_1N_2-N_2^2$&$\text{Sum-DoF}\geq \min\{\lfloor \frac{R+N_1^2+N_2^2}{N_1+N_2} \rfloor,M_1+N_2,M_1+M_2\}$\\
			\cline{4-5}
			 &\multicolumn{2}{c}{\hspace{4.4cm}{\footnotesize \text{Case-3}}}\vline&$R< N_1N_2-N_2^2$ &$\text{Sum-DoF}\geq \min\{N_2+\lfloor\frac{R}{N_2}\rfloor,M_1+N_2,M_1+M_2\}$\\
			\hline
		\end{tabular}
	\end{table*}
	%\vspace{-0.08in}
	%	\begin{table}
		%		\centering
		%		\renewcommand\arraystretch{1.8}
		%		\caption{Required Minimum Number of RIS Elements}
		%		\label{MinR}
		%		\begin{tabular}{|c|c|}
			%			\hline
			%		\text{Case 1} & $R\geq M_2$,\ ($M_2 < N_1+N_2$)\\
			%			\hline
			%			\text{Case 2} & $R\geq N_1$\\
			%			\hline
			%			\text{Case 3} & $R\geq N_2$,\ ($N_2 < M_1+M_2$)\\
			%			\hline
			%			\text{Case 4} & $R\geq M_1$\\
			%			\hline
			%		\end{tabular}
		%	\end{table}
	\section{Proposed Achievable Sum-DoF}\label{proof}
	%	In order to achieve Table I, we propose a general framework of achievable scheme, in which we first cancell the interference by RIS beamforming and then remove the residual interference by transmit beamforming. Below, we present the specific achievable schemes case by case. 
	
		Similar to \cite{jafar}, due to $\max\{M_1,N_1\}\geq  \max\{M_2,N_2\}$,  we can divide all antenna configurations into 3 cases (please refer to Table I for the case division and the achievable sum-DoF), and present the achievable schemes of 3 cases below.  Note that the union of 3 cases includes all antenna configurations satisfying $\max\{M_1,N_1\}\geq  \max\{M_2,N_2\}$.

	\subsection{ Case-1 (Row-Row Elimination): $M_1 \geq \{M_2,N_1,N_2\}$ and $M_2 \ge N_1$}

	First, the active RIS is used to cancel the interference links, i.e., $\mathbf{H}_{21} \in \mathbb{C}^{N_1 \times M_2},\mathbf{H}_{12} \in \mathbb{C}^{N_2 \times M_1}$. Due to $M_2 \ge N_1$ and $M_1 \ge N_2$, to reduce the rank of $\mathbf{H}_{21},\mathbf{H}_{12}$, we thus eliminate the first $f_1 \in [0,N_1]$ \textit{rows} of $\mathbf{H}_{21}$ ($f_1$ interference links) and the first $f_2\in [0,N_2]$ \textit{rows} of $\mathbf{H}_{12}$ by RIS beamforming. This is why we call it row-row elimination. This is done by solving the following set of equations 
	\begin{equation}
		\begin{cases}
			\mathbf{H}_{21}  + \mathbf{G}_1 {\bm \Psi} \mathbf{D}_2 = \begin{bmatrix}
				\mathbf{0}_{f_1 \times M_2}\\
				\widetilde{\mathbf{H}}_{21}
			\end{bmatrix},\\	  
			\mathbf{H}_{12}  + \mathbf{G}_2 {\bm \Psi} \mathbf{D}_1 = \begin{bmatrix}
				\mathbf{0}_{f_2 \times M_1} \\
				\widetilde{\mathbf{H}}_{12}
			\end{bmatrix},
		\end{cases}
		\label{C1-2}
	\end{equation}
	where $\widetilde{\mathbf{H}}_{21} \in \mathbb{C}^{(N_1-f_1) \times M_2}$ and $\widetilde{\mathbf{H}}_{12} \in \mathbb{C}^{(N_2-f_2) \times M_1}$ are arbitrary matrices with full-rank. In order to solve \eqref{C1-2}, we can vectorize it as 
	\begin{equation}
		\begin{cases}
			\mathbf{h}_{21} +	{\bm{\Gamma}}_1 {\bm {\psi}} = \begin{bmatrix}
				\mathbf{0}_{f_1M_2 \times 1} \\ \widetilde{\mathbf{h}}_{21} 
			\end{bmatrix}, \\
			\mathbf{h}_{12} +	{\bm{\Gamma}}_2 {\bm {\psi}} = \begin{bmatrix}
				\mathbf{0}_{f_2M_1  \times 1} \\ \widetilde{\mathbf{h}}_{12} 
			\end{bmatrix}, 
		\end{cases}\label{C1-4}
	\end{equation}
	where ${\bm \psi} \in \mathbb{C}^{R \times 1} \triangleq [\psi_1,\psi_2,...,\psi_R]^T$, $\mathbf{h}_{21} \in \mathbb{C}^{N_1M_2 \times 1} \triangleq \text{vec}(\mathbf{H}_{21}^T)$, ${\bm{\Gamma}}_1 \in \mathbb{C}^{N_1M_2 \times R} \triangleq [{\mathbf{V}_1^{1}}^T,{\mathbf{V}_2^{1}}^T,...,{\mathbf{V}_{N_1}^{1}}^T]^T$ with $[\mathbf{V}_k^{1}]^{i,j} \triangleq [\mathbf{G}_1]^{k,j} [\mathbf{D}_2]^{j,i} $, $\widetilde{\mathbf{h}}_{21} \in \mathbb{C}^{(N_1-f_1)M_2 \times 1} \triangleq \text{vec}(\widetilde{\mathbf{H}}^{T}_{21})$, $\mathbf{h}_{12} \in \mathbb{C}^{N_2M_1 \times 1} \triangleq \text{vec}(\mathbf{H}_{12}^T)$, ${\bm{\Gamma}}_2 \in \mathbb{C}^{N_2M_1 \times R} \triangleq [{\mathbf{V}_1^{2}}^T,{\mathbf{V}_2^{2}}^T,...,{\mathbf{V}_{N_2}^{2}}^T]^T$ with $[\mathbf{V}_k^{2}]^{i,j} \triangleq [\mathbf{G}_2]^{k,j} [\mathbf{D}_1]^{j,i} $, $\widetilde{\mathbf{h}}_{12} \in \mathbb{C}^{(N_2-f_2)M_1 \times 1}  \triangleq \text{vec}(\widetilde{\mathbf{H}}^{T}_{12})$. Since we require the specific outputs (interference cancellation) at the first $f_1M_2$ rows in the first equation of \eqref{C1-4} and the first $f_2M_1$ rows the second equation of  \eqref{C1-4}, the non-zero solutions (RIS beamforming) of \eqref{C1-4} exists if $f_1M_2 + f_2M_1 \le R$. By this way, the rank of $\mathbf{H}_{21}$ reduces to $N_1-f_1$ and the rank of $\mathbf{H}_{12}$ reduces to $N_2-f_2$. We further denote the equivalent matrices after the RIS beamforming by $\overline{\mathbf{H}}_{11} \triangleq \mathbf{H}_{11} + \mathbf{G}_1 {\bm \Psi} \mathbf{D}_1$, $\overline{\mathbf{H}}_{21} \triangleq [\mathbf{0}_{M_2 \times f_1}, \widetilde{\mathbf{H}}^T_{21}]^T$, $\overline{\mathbf{H}}_{22} \triangleq \mathbf{H}_{22} + \mathbf{G}_2 {\bm \Psi} \mathbf{D}_2$, and $\overline{\mathbf{H}}_{12} \triangleq [\mathbf{0}_{M_1 \times f_2}, \widetilde{\mathbf{H}}^T_{12}]^T$.
	
	In the following, by adopting the zero-forcing and interference decoding in \cite{jafar}, we aim at eliminating the residual interference after RIS beamforming. The input-output relationships after the aforementioned RIS beamforming are given as %
	\begin{eqnarray}
		&& \mathbf{y}_1 = \overline{\mathbf{H}}_{11}\mathbf{P}_1\mathbf{x}_1 + \overline{\mathbf{H}}_{21}\mathbf{P}_2\mathbf{x}_2+\mathbf{z}_1, \label{A1} \\
		&& \mathbf{y}_2 = \overline{\mathbf{H}}_{22}\mathbf{P}_2\mathbf{x}_2 + \overline{\mathbf{H}}_{12}\mathbf{P}_1\mathbf{x}_1+\mathbf{z}_2. \label{A2}
	\end{eqnarray} 

		Second, we consider the transmit zero-forcing design. The space for zero-forcing at $\text{Rx}_1$ and $\text{Rx}_2$ is $\min\{M_1-(N_2-f_2),N_1\}$ and $\min\{M_2-(N_1-f_1),N_2\}$, respectively. After that, the remaining vector space at $\text{Rx}_1$ and $\text{Rx}_2$ is $\max\{N_1+N_2-M_1-f_2,0\}$ and $\max\{N_1+N_2-M_2-f_1,0\}$, respectively. Finally, we utilize the interference decoding, and send symbols in the vector space of $\min\{\max\{N_1+N_2-M_1-f_2,0\},\max\{N_1+N_2-M_2-f_1,0\}\}$ with the corresponding transmitter. Thus, the created interference at another receiver can be decoded. Then, we divide Case-1 into four sub-cases to analyze the achievable sum-DoF.
	\begin{enumerate}
		\item If $M_1-(N_2-f_2)\le N_1$ and $M_2-(N_1-f_1)\le N_2$, which indicates that there is remaining space for interference decoding. In this sub-case, the achievable sum-DoF is $M_1 - (N_2 - f_2)+M_2 - (N_1 - f_1)+\min\{N_1 + N_2 - M_1 - f_2, N_1 + N_2 - M_2 - f_1\}=\min\{M_2+f_1,M_1+f_2\}$. Since  $M_1-(N_2-f_2)\le N_1$, we know $M_1+f_2\le N_1+N_2$. Similarly, we know $M_2+f_1\le N_1+N_2$. Finally, we can express the achievable sum-DoF as $\min\{M_2+f_1,M_1+f_2,N_1+N_2,M_1+M_2\}$ by further considering the limitation of transmitter space $M_1+M_2$.
		\item If $M_1-(N_2-f_2)> N_1$ and $M_2-(N_1-f_1)> N_2$, which indicates that the sum-DoF is restricted to the number of receive antennas. In this sub-case, the achievable sum-DoF is $N_1+N_2+0=N_1+N_2$. Since $M_1-(N_2-f_2)>N_1$, we know $M_1+f_2>N_1+N_2$. Similarly, we know  $M_2+f_1>N_1+N_2$. Finally, we can express the achievable sum-DoF as $\min\{M_2+f_1,M_1+f_2,N_1+N_2,M_1+M_2\}$ by further considering the limitation of transmitter space $M_1+M_2$.
		\item If $M_1-(N_2-f_2)\le N_1$ and $M_2-(N_1-f_1)> N_2$, which indicates that no remaining space can be utilized for interference decoding. In this sub-case, the achievable sum-DoF is $M_1-(N_2-f_2)+N_2+0=M_1+f_2$. Since $M_1-(N_2-f_2)\le N_1$, we know $N_1+N_2\ge M_1+f_2$. Similarly, we know that $M_2+f_1>N_1+N_2$ thus $M_2+f_1>M_1+f_2$. Finally, we can express the achievable sum-DoF as $\min\{M_2+f_1,M_1+f_2,N_1+N_2,M_1+M_2\}$ by further considering the limitation of transmitter space $M_1+M_2$.
		\item If $M_1-(N_2-f_2)> N_1$ and $M_2-(N_1-f_1)\le N_2$, which indicates that no remaining space can be utilized for interference decoding. In this sub-case, the achievable sum-DoF is $N_1+M_2-(N_1-f_1)+0=M_2+f_1$. Since $M_2-(N_1-f_1)\le N_2$, we know $N_1+N_2\ge M_2+f_1$. Similarly, we know that $M_1+f_2>N_1+N_2$ thus $M_1+f_2>M_2+f_1$. Finally, we can express the achievable sum-DoF as $\min\{M_2+f_1,M_1+f_2,N_1+N_2,M_1+M_2\}$ by further considering the limitation of transmitter space $M_1+M_2$.
	\end{enumerate}

	 Since $M_1+M_2\ge \{M_2+f_1,N_1+N_2\}$ in this case, we can simplify the expression of the achievable sum-DoF as $\min\{M_2+f_1,M_1+f_2,N_1+N_2\}$. 
	
	An integer optimization problem for optimal $f_1^*$ and $f_2^*$ is formulated as 
	\begin{eqnarray}
		\mathcal{P}_1:	\max_{f_1,f_2 \in \mathbb{N}} && \!\!\!\!\!\! \min\{M_2+f_1,M_1+f_2,N_1+N_2\}   \nonumber \\
		\text{s.t.} && \!\!\!\!\!\! f_1M_2 + f_2M_1 \le R, f_i \in [0,N_i],i=1,2. \nonumber 
	\end{eqnarray} When $R\ge M_1M_2-M_2^2$, due to $\min$ function, once if $M_2+f_1=M_1+f_2$, the objective reaches the maximum. 
	Specifically, by exhausting $R$, we aim to solve the following two equations
	\begin{equation}
		\begin{cases}
			M_2+f_1 = M_1+f_2,\\
			f_1M_2+f_2M_1=R.
		\end{cases} \label{Case1-E1}
	\end{equation}
	By solving \eqref{Case1-E1} with $f_1 \in [0,N_1],\,f_2 \in [0,N_2], \, f_1,f_2\in \mathbb{N}$, the optimal $f_1^*$ and $f_2^*$ are given by
	\begin{equation}
		\label{Case1-E2}
		(f_1^*,f_2^*)=
		( \min \{\lfloor \gamma_1 \rfloor,N_1\}, \min\{\lfloor \gamma_2 \rfloor,N_2\}),
	\end{equation}
	where $\gamma_1  =  \frac{R-M_1M_2+M_1^2}{M_1+M_2} $ and $  \gamma_2  = \frac{R-M_1M_2+M_2^2}{M_1+M_2} $. Substituting \eqref{Case1-E2} into the objective function of Problem $	\mathcal{P}_1$ gives $\min\{\lfloor \frac{R+M_1^2+M_2^2}{M_1+M_2} \rfloor,M_2+N_1,N_1+N_2\}$.
 When $R< M_1M_2-M_2^2$, however, this equivalence relationship, i.e., $M_2+f_1=M_1+f_2$, cannot hold.  Due to $M_1 > M_2$, RIS is devoted to elevate bottleneck term $M_2+f_1$ in $\min$ function, leading to
	\begin{equation}
		\begin{cases}		 
			f_1M_2+f_2M_1=R, \\
			M_2+f_1 < M_1+f_2, \\
			f_2 = 0.
		\end{cases}		
		\label{Case1-E3}
	\end{equation}
	By solving \eqref{Case1-E3} with $f_1 \in [0,N_1],\,f_2 \in [0,N_2], \, f_1,f_2\in \mathbb{N}$, the optimal $f_1^*$ and $f_2^*$ are given by
	\begin{equation}
		(f_1^*,f_2^*)=
		(\min \{\lfloor R/M_2 \rfloor, N_1\},0).\label{Case1-E4}
	\end{equation}
	Substituting \eqref{Case1-E4} into the objective function of Problem $	\mathcal{P}_1$ gives $\min\{M_2+\lfloor R/M_2 \rfloor,M_2+N_1,N_1+N_2\}$.
	Therefore, it can be seen that Table I Case-1 is achieved, where $R = M_1M_2-M_2^2$ satisfies $\lfloor \frac{R+M_1^2+M_2^2}{M_1+M_2} \rfloor = M_2+\lfloor R/M_2 \rfloor$.
	%By SVD, i.e., $\mathbf{H} = \mathbf{U}{\bm {\Lambda} }\mathbf{V}^H$, where $\mathbf{U}$ and $\mathbf{V}$ are unitary matrices and ${\bm {\Lambda} }$ is a block-diagonal-wise matrix with singular values, we can represent the input-output relationship as
	%\begin{eqnarray}
	%	&& \mathbf{y}'_1 = \mathbf{H}_{11}'\mathbf{x}_1 + {\bm {\Lambda} }_{21} \mathbf{x}'_2, \\
	%	&& \mathbf{y}'_2 = \mathbf{H}_{22}'\mathbf{x}_2 + {\bm {\Lambda} }_{12} \mathbf{x}'_1,
	%\end{eqnarray}
	%where $\mathbf{y}'_1 = \mathbf{U}_{21}^H\mathbf{y}_1$, $\mathbf{H}_{11}'=\mathbf{U}_{21}^H\overline{\mathbf{H}}_{11}\mathbf{P}_{1}$, $\mathbf{x}'_2 = \mathbf{V}_{21}^H\mathbf{P}_2\mathbf{x}_2$, $\mathbf{y}'_2 = \mathbf{U}_{12}^H\mathbf{y}_2$, $\mathbf{H}_{22}'=\mathbf{U}_{12}^H\overline{\mathbf{H}}_{22}\mathbf{P}_{2}$, and $\mathbf{x}'_1 = \mathbf{V}_{12}^H\mathbf{P}_1\mathbf{x}_1$. It suffices to zero-force the interference terms ${\bm {\Lambda} }_{21} \mathbf{x}'_2$ and ${\bm {\Lambda} }_{12} \mathbf{x}'_1$.
	
 \subsection{Case-2 (Column-Row Elimination): $M_1 \geq N_2$ and $N_1 \ge M_2$}

	First, the active RIS is used to cancel the interference links, i.e., $\mathbf{H}_{21} \in \mathbb{C}^{N_1 \times M_2},\mathbf{H}_{12} \in \mathbb{C}^{N_2 \times M_1}$. Due to $N_1 \ge M_2$ and $M_1 \geq N_2$, to reduce the rank of $\mathbf{H}_{21}, \mathbf{H}_{12}$, we thus eliminate the first $f_1 \in [0,M_2]$ \textit{columns} of $\mathbf{H}_{21}$ and the first $f_2 \in [0,N_2]$ \textit{rows} of $\mathbf{H}_{12}$ by RIS beamforming.   This is why we call it column-row elimination. This is done by solving the following set of equations 
	\begin{equation}
		\begin{cases}
			\mathbf{H}_{21}  + \mathbf{G}_1 {\bm \Psi} \mathbf{D}_2 = \begin{bmatrix}
				\mathbf{0}_{N_1 \times f_1},
				\widetilde{\mathbf{H}}_{21}
			\end{bmatrix},  \\	 
			\mathbf{H}_{12}  + \mathbf{G}_2 {\bm \Psi} \mathbf{D}_1 = \begin{bmatrix}
				\mathbf{0}_{f_2 \times M_1} \\
				\widetilde{\mathbf{H}}_{12}
			\end{bmatrix}, 
		\end{cases}		\label{C2-2}
	\end{equation}
	respectively, where $\widetilde{\mathbf{H}}_{21} \in \mathbb{C}^{N_1 \times (M_2-f_1)}$ and $\widetilde{\mathbf{H}}_{12} \in \mathbb{C}^{(N_2-f_2) \times M_1}$ are arbitrary matrices with full-rank. To solve \eqref{C2-2}, we can vectorize them as 
	\begin{equation}
		\begin{cases}
			\mathbf{h}_{21} +	{\bm{\Gamma}}_1 {\bm {\psi}} = \begin{bmatrix}
				\mathbf{0}_{f_1N_1 \times 1} \\ \widetilde{\mathbf{h}}_{21} 
			\end{bmatrix}, \\
			\mathbf{h}_{12} +	{\bm{\Gamma}}_2 {\bm {\psi}} = \begin{bmatrix}
				\mathbf{0}_{f_2M_1  \times 1} \\ \widetilde{\mathbf{h}}_{12} 
			\end{bmatrix},
		\end{cases}
		\label{C2-4}
	\end{equation}
	where ${\bm \psi} \in \mathbb{C}^{R \times 1} \triangleq [\psi_1,\psi_2,...,\psi_R]^T$, $\mathbf{h}_{21} \in \mathbb{C}^{N_1M_2 \times 1} \triangleq \text{vec}(\mathbf{H}_{21})$, ${\bm{\Gamma}}_1 \in \mathbb{C}^{N_1M_2 \times R} \triangleq [{\mathbf{V}_1^{1}}^T,{\mathbf{V}_2^{1}}^T,...,{\mathbf{V}_{M_2}^{1}}^T]^T$ with $[\mathbf{V}_k^{1}]^{i,j} \triangleq [\mathbf{G}_1]^{i,j} [\mathbf{D}_2]^{j,k} $, $\widetilde{\mathbf{h}}_{21} \in \mathbb{C}^{(M_2-f_1)N_1 \times 1} \triangleq \text{vec}(\widetilde{\mathbf{H}}_{21})$, $\mathbf{h}_{12} \in \mathbb{C}^{N_2M_1 \times 1} \triangleq \text{vec}(\mathbf{H}_{12}^T)$, ${\bm{\Gamma}}_2 \in \mathbb{C}^{N_2M_1 \times R} \triangleq [{\mathbf{V}_1^{2}}^T,{\mathbf{V}_2^{2}}^T,...,{\mathbf{V}_{N_2}^{2}}^T]^T$ with $[\mathbf{V}_k^{2}]^{i,j} \triangleq [\mathbf{G}_2]^{k,j} [\mathbf{D}_1]^{j,i} $, $\widetilde{\mathbf{h}}_{12} \in \mathbb{C}^{(N_2-f_2)M_1 \times 1}  \triangleq \text{vec}(\widetilde{\mathbf{H}}^T_{12})$. Since we require the specific outputs (interference cancellation) at the first $f_1N_1$ rows of the first equation of \eqref{C2-4} and the first $f_2M_1$ rows of the second equation of \eqref{C2-4}, the non-zero solutions (RIS beamforming) of \eqref{C2-4} exists if $f_1N_1 + f_2M_1 \le R$. By this way, the rank of $\mathbf{H}_{21}$ reduces to $M_2-f_1$ and the rank of $\mathbf{H}_{12}$ reduces to $N_2-f_2$. We further denote the equivalent matrices after the RIS beamforming by $\overline{\mathbf{H}}_{11} \triangleq \mathbf{H}_{11} + \mathbf{G}_1 {\bm \Psi} \mathbf{D}_1$, $\overline{\mathbf{H}}_{21} \triangleq [\mathbf{0}_{N_1 \times f_1}, \widetilde{\mathbf{H}}_{21}]$, $\overline{\mathbf{H}}_{22} \triangleq \mathbf{H}_{22} + \mathbf{G}_2 {\bm \Psi} \mathbf{D}_2$, $\overline{\mathbf{H}}_{12} \triangleq [\mathbf{0}_{M_1 \times f_2}, \widetilde{\mathbf{H}}^T_{12}]^T$. The input-output relationships after the above RIS beamforming are the same as \eqref{A1} and \eqref{A2}.
	
	%The input-output relationship after the aforementioned RIS beamforming is given as follows.
	%\begin{eqnarray}
	%&& \mathbf{y}_1 = \overline{\mathbf{H}}_{11}\mathbf{P}_1\mathbf{x}_1 + \overline{\mathbf{H}}_{21}\mathbf{P}_2\mathbf{x}_2, \\
	%&& \mathbf{y}_2 = \overline{\mathbf{H}}_{22}\mathbf{P}_2\mathbf{x}_2 + \overline{\mathbf{H}}_{12}\mathbf{P}_1\mathbf{x}_1.
	%\end{eqnarray}
	Second, we consider the zero-forcing design. The space for zero-forcing at $\text{Rx}_1$ and $\text{Rx}_2$ is $\min\{M_1-(N_2-f_2),N_1\}$ and $\min\{f_1,N_2\}$, respectively. After that, the remaining vector space at $\text{Rx}_1$ and $\text{Rx}_2$ is $\max\{N_1+N_2-M_1-f_2,0\}$ and $\max\{N_2-f_1,0\}$, respectively. Finally, we utilize the interference decoding, and send symbols in the vector space of $\min\{\max\{N_1+N_2-M_1-f_2,0\},\max\{N_2-f_1,0\}\}$ with the corresponding transmitter. Thus, the created interference at another receiver can be decoded. Then, we divide Case-2 into four sub-cases to analyze the achievable sum-DoF. 
	\begin{enumerate}
		\item If $M_1-(N_2-f_2)\le N_1$ and $f_1\le N_2$, which indicates that there is remaining space for interference decoding. By adding the spaces up, we obtain $M_1 - (N_2 - f_2)+f_1+\min\{N_2 - f_1, N_1 + N_2 - M_1 - f_2\}=\min\{N_1+f_1,M_1+f_2\}$. Since, $M_1-(N_2-f_2)\le N_1$, we know $M_1+f_2\le N_1+N_2$. Finally, we can express the achievable sum-DoF as $\min\{N_1+f_1,M_1+f_2,M_1+M_2, N_1+N_2\}$ by further considering the limitation of transmitter space $M_1+M_2$.
		\item If $M_1-(N_2-f_2)> N_1$ and $f_1> N_2$, which indicates that the sum-DoF is restricted to the number of receive antennas. In this sub-case, the achievable sum-DoF is $N_1+N_2+0=N_1+N_2$. Since $M_1-(N_2-f_2)>N_1$, we know that $M_1+f_2>N_1+N_2$. Similarly, we know that $N_1+f_1>N_1+N_2$. Finally, we can express the achievable sum-DoF as $\min\{N_1+f_1,M_1+f_2,N_1+N_2,M_1+M_2\}$ by further considering the limitation of transmitter space $M_1+M_2$.
		\item If $M_1-(N_2-f_2)\le N_1$ and $f_1> N_2$, which indicates that no remaining space can be utilized for interference decoding. In this sub-case, the achievable sum-DoF is $M_1-(N_2-f_2)+N_2+0=M_1+f_2$. Since $M_1-(N_2-f_2)\le N_1$, we know that $N_1+N_2\ge M_1+f_2$. Similarly, we know that $N_1+f_1>N_1+N_2$ thus $N_1+f_1>M_1+f_2$. Finally, we can express the achievable sum-DoF as $\min\{N_1+f_1,M_1+f_2,N_1+N_2,M_1+M_2\}$ by further considering the limitation of transmitter space $M_1+M_2$.
		\item If $M_1-(N_2-f_2)> N_1$ and $f_1\le N_2$, which indicates that no remaining space can be utilized for interference decoding. In this sub-case, the achievable sum-DoF is $N_1+f_1+0=N_1+f_1$. Since $f_1\le N_2$, we know that $N_1+N_2 \ge N_1+f_1$. Similarly, we know that $M_1+f_2>N_1+N_2$ thus $M_1+f_2>N_1+f_1$. Finally, we can express the achievable sum-DoF as $\min\{N_1+f_1,M_1+f_2,N_1+N_2,M_1+M_2\}$ by further considering the limitation of transmitter space $M_1+M_2$.
	\end{enumerate}

	An integer optimization problem for optimal $f_1^*$ and $f_2^*$ is formulated as%given by
	\begin{eqnarray}
		\mathcal{P}_2:	\max_{f_1,f_2 \in \mathbb{N}} && \!\!\!\!\! \!\!\!\!\! \min\{N_1+f_1,M_1+f_2,M_1+M_2,N_1+N_2\} \nonumber   \\
		\text{s.t.} && \!\!\!\!\! \!\!\!\!\! f_1N_1 + f_2M_1 \le R, f_1\in[0,M_2],f_2\in[0,N_2]. \nonumber 
	\end{eqnarray} 
  When $M_1\geq N_1$ and $R\geq M_1N_1-N_1^2$, or $N_1>M_1$ and $R\geq M_1N_1-M_1^2$, due to $\min$ function, once if $N_1+f_1=M_1+f_2$, the objective reaches the maximum. 
	Specifically, by exhausting $R$, we aim to solve the following two equations
	\begin{equation}
		\begin{cases}
			N_1+f_1 = M_1+f_2,\\
			f_1N_1+f_2M_1=R.
		\end{cases} \label{Case2-E1}
	\end{equation}
	By solving \eqref{Case2-E1} with $f_1 \in [0,M_2],\,f_2 \in [0,N_2], \, f_1,f_2\in \mathbb{N}$, the optimal $f_1^*$ and $f_2^*$ are given by
	\begin{equation}
		(f_1^*,f_2^*)=( \min \{\lfloor \gamma_3 \rfloor,M_2\}, \min\{\lfloor \gamma_4 \rfloor,N_2\}), \label{Case2-E2}
	\end{equation}
	where $\gamma_3  =  \frac{R-M_1N_1+M_1^2}{M_1+N_1} $ and $  \gamma_4  = \frac{R-M_1N_1+N_1^2}{M_1+N_1} $. Substituting \eqref{Case2-E2} into the objective function of Problem $	\mathcal{P}_2$ gives $\min\{\lfloor \frac{R+M_1^2+N_1^2}{M_1+N_1} \rfloor,M_1+N_2, M_2+N_1, M_1+M_2, N_1+N_2\}$. To characterize the achievable sum-DoF more precisely, we divide Case-2 into two sub-cases: Case-2-1 and Case-2-2.  
	
	\subsubsection{Case-2-1 ($M_1\geq N_1$)}
	When $M_1\geq N_1$, the achievable sum-DoF derived above is simplified as $\min\{\lfloor \frac{R+M_1^2+N_1^2}{M_1+N_1} \rfloor,M_2+N_1, N_1+N_2\}$ since $M_1+N_2\geq N_1+N_2$ and $M_1+M_2\geq M_2+N_1$.
	When $R< M_1N_1-N_1^2$, however, this equivalence relationship, i.e., $N_1+f_1=M_1+f_2$, cannot hold. Due to $M_1 > N_1$, RIS is devoted to elevate bottleneck term $N_1+f_1$ in $\min$ function, leading to
	\begin{equation}
		\begin{cases}		 
			f_1N_1+f_2M_1=R, \\		
			N_1 + f_1 < M_1 + f_2 \\
			f_2 = 0, \\
		\end{cases}		
		\label{Case2-E3}
	\end{equation}
	By solving \eqref{Case2-E3} with $f_1 \in [0,M_2],\,f_2 \in [0,N_2], \, f_1,f_2\in \mathbb{N}$, the optimal $f_1^*$ and $f_2^*$ are given by
	\begin{equation}
		(f_1^*,f_2^*)=\left(\min \left\{\left\lfloor R/N_1\right\rfloor, M_2\right\},0\right). \label{Case2-E4}
	\end{equation}
	Substituting \eqref{Case2-E4} into the objective function of Problem $	\mathcal{P}_2$ gives $\min\{N_1+\lfloor R/N_1 \rfloor,M_2+N_1,N_1+N_2\}$.
	Therefore, it can be seen that Table I Case-2-1 is achieved, where $R = M_1N_1-N_1^2$ satisfies $\lfloor \frac{R+M_1^2+N_1^2}{M_1+N_1} \rfloor = N_1+\lfloor R/N_1 \rfloor$.
	
	\subsubsection{Case-2-2 ($N_1>M_1$)}
	When $N_1>M_1$, the achievable sum-DoF derived above is simplified as $\min\{\lfloor \frac{R+M_1^2+N_1^2}{M_1+N_1} \rfloor,M_1+N_2, M_1+M_2\}$ since $N_1+N_2>M_1+N_2$ and $M_2+N_1>M_1+M_2$. When $R< M_1N_1-M_1^2$, however, this equivalence relationship, i.e., $N_1+f_1=M_1+f_2$, cannot hold.
	Due to $N_1 > M_1$, RIS is devoted to elevate bottleneck term $M_1+f_2$ in $\min$ function, leading to
	\begin{equation}
		\begin{cases}		 
			f_1N_1+f_2M_1=R, \\		
			N_1 + f_1 > M_1 + f_2 \\
			f_1 = 0, \\
		\end{cases}		
		\label{Case4-E3}
	\end{equation}
	By solving \eqref{Case4-E3} with $f_1 \in [0,M_2],\,f_2 \in [0,N_2], \, f_1,f_2\in \mathbb{N}$, the optimal $f_1^*$ and $f_2^*$ are given by
	\begin{equation}
		(f_1^*,f_2^*)=\left(0,\min \{\lfloor R/M_1\rfloor, N_2\}\right). \label{Case4-E4}
	\end{equation}
	Substituting \eqref{Case4-E4} into the objective function of Problem $	\mathcal{P}_2$ gives $\min\{M_1+\lfloor R/M_1 \rfloor,M_1+N_2,M_1+M_2\}$.
	Therefore, it can be seen that Table I Case-2-2 is achieved, where $R =M_1N_1-M_1^2$ satisfies $\lfloor \frac{R+M_1^2+N_1^2}{M_1+N_1} \rfloor = M_1+\lfloor R/M_1 \rfloor$.

		\subsection{ Case-3 (Column-Column Elimination): $N_1 \ge \{M_1,M_2,N_2\}$ and $N_2 \ge M_1$} 
	
	First, the active RIS is used to cancel the interference links, i.e., $\mathbf{H}_{21} \in \mathbb{C}^{N_1 \times M_2},\mathbf{H}_{12} \in \mathbb{C}^{N_2 \times M_1}$. Due to $N_1 \geq M_2$ and $N_2 \geq M_1$, to reduce the rank of $\mathbf{H}_{21}, \mathbf{H}_{12}$, we thus eliminate the first $f_1 \in [0,M_2]$ \textit{columns} of $\mathbf{H}_{21}$ and the first $f_2 \in [0,M_1]$ \textit{columns} of $\mathbf{H}_{12}$ by RIS beamforming.  This is why we call it column-column elimination. This is done by solving the following set of equations 
	\begin{equation}
		\begin{cases}
			\mathbf{H}_{21}  + \mathbf{G}_1 {\bm \Psi} \mathbf{D}_2 = \begin{bmatrix}
				\mathbf{0}_{N_1 \times f_1},
				\widetilde{\mathbf{H}}_{21}
			\end{bmatrix},\\
			\mathbf{H}_{12}  + \mathbf{G}_2 {\bm \Psi} \mathbf{D}_1 = \begin{bmatrix}
				\mathbf{0}_{N_2 \times f_2},
				\widetilde{\mathbf{H}}_{12}
			\end{bmatrix},
		\end{cases}			\label{C3-2} 
	\end{equation}
	respectively, where $\widetilde{\mathbf{H}}_{21} \in \mathbb{C}^{N_1 \times (M_2-f_1)}$ and $\widetilde{\mathbf{H}}_{12} \in \mathbb{C}^{N_2 \times (M_1-f_2)}$ are arbitrary matrices with full-rank. To solve \eqref{C3-2}, we can vectorize them as 
	\begin{equation}
		\begin{cases}
			\mathbf{h}_{21} +	{\bm{\Gamma}}_1 {\bm {\psi}} = \begin{bmatrix}
				\mathbf{0}_{f_1N_1 \times 1} \\ \widetilde{\mathbf{h}}_{21} 
			\end{bmatrix}, \\
			\mathbf{h}_{12} +	{\bm{\Gamma}}_2 {\bm {\psi}} = \begin{bmatrix}
				\mathbf{0}_{f_2N_2  \times 1} \\ \widetilde{\mathbf{h}}_{12} 
			\end{bmatrix}, 
		\end{cases}\label{C3-4}
	\end{equation}
	where ${\bm \psi} \in \mathbb{C}^{R \times 1} \triangleq [\psi_1,\psi_2,\cdots,\psi_R]^T$, $\mathbf{h}_{21} \in \mathbb{C}^{N_1M_2 \times 1} \triangleq \text{vec}(\mathbf{H}_{21})$, ${\bm{\Gamma}}_1 \in \mathbb{C}^{N_1M_2 \times R} \triangleq [{\mathbf{V}_1^{1}}^T,{\mathbf{V}_2^{1}}^T,\cdots,{\mathbf{V}_{M_2}^{1}}^T]^T$ with $[\mathbf{V}_k^{1}]^{i,j} \triangleq [\mathbf{G}_1]^{i,j} [\mathbf{D}_2]^{j,k} $, $\widetilde{\mathbf{h}}_{21} \in \mathbb{C}^{(M_2-f_1)N_1 \times 1} \triangleq \text{vec}(\widetilde{\mathbf{H}}_{21})$, $\mathbf{h}_{12} \in \mathbb{C}^{N_2M_1 \times 1} \triangleq \text{vec}(\mathbf{H}_{12})$, ${\bm{\Gamma}}_2 \in \mathbb{C}^{N_2M_1 \times R} \triangleq [{\mathbf{V}_1^{2}}^T,{\mathbf{V}_2^{2}}^T,\cdots,{\mathbf{V}_{M_1}^{2}}^T]^T$ with $[\mathbf{V}_k^{2}]^{i,j} \triangleq [\mathbf{G}_2]^{i,j} [\mathbf{D}_1]^{j,k} $, $\widetilde{\mathbf{h}}_{12} \in \mathbb{C}^{(M_1-f_2)N_2 \times 1} \triangleq \text{vec}(\widetilde{\mathbf{H}}_{12})$. Since we require the specific outputs (interference cancellation) at the first $f_1N_1$ rows of the first equation of \eqref{C3-4} and the first $f_2N_2$ rows of the second equation of \eqref{C3-4}, the non-zero solutions (RIS beamforming) of \eqref{C3-4} exists if $f_1N_1 + f_2N_2 \le R$. By this way, the rank of $\mathbf{H}_{21}$ reduces to $M_2-f_1$ and the rank of $\mathbf{H}_{12}$ reduces to $M_1-f_2$. We further denote the equivalent matrices after the RIS beamforming by $\overline{\mathbf{H}}_{11} \triangleq \mathbf{H}_{11} + \mathbf{G}_1 {\bm \Psi} \mathbf{D}_1$, $\overline{\mathbf{H}}_{21} \triangleq [\mathbf{0}_{N_1 \times f_1}, \widetilde{\mathbf{H}}_{21}]$, $\overline{\mathbf{H}}_{22} \triangleq \mathbf{H}_{22} + \mathbf{G}_2 {\bm \Psi} \mathbf{D}_2$, $\overline{\mathbf{H}}_{12} \triangleq [\mathbf{0}_{N_2 \times f_2}, \widetilde{\mathbf{H}}_{12}]$. The input-output relationships after the above RIS beamforming are the same as \eqref{A1} and \eqref{A2}.
	
	%The input-output relationship after the aforementioned RIS beamforming is given as follows.
	%\begin{eqnarray}
	%&& \mathbf{y}_1 = \overline{\mathbf{H}}_{11}\mathbf{P}_1\mathbf{x}_1 + \overline{\mathbf{H}}_{21}\mathbf{P}_2\mathbf{x}_2, \\
	%&& \mathbf{y}_2 = \overline{\mathbf{H}}_{22}\mathbf{P}_2\mathbf{x}_2 + \overline{\mathbf{H}}_{12}\mathbf{P}_1\mathbf{x}_1.
	%\end{eqnarray}
	Second, we consider the zero-forcing design. The space for zero-forcing at $\text{Rx}_1$ and $\text{Rx}_2$ is $f_2$ and $\min\{f_1,N_2\}$, respectively. After that, the remaining vector space at $\text{Rx}_1$ and $\text{Rx}_2$ is $N_1-f_2$ and $\max\{N_2-f_1,0\}$, respectively. Finally, we utilize the interference decoding, and send symbols in the vector space of $\min\{N_1-f_2,\max\{N_2-f_1,0\}\}$ with the corresponding transmitter. Thus, the created interference at another receiver can be decoded. Then, we divide Case-3 into two sub-cases to analyze the achievable sum-DoF. 
	\begin{enumerate}
		\item If $f_1\le N_2$, which indicates that there is remaining space for interference decoding. In this sub-case, the achievable sum-DoF is $f_2+f_1+\min\{N_1 - f_2, N_2 - f_1\}=\min\{N_1+f_1,N_2+f_2\}$. Since $f_1\le N_2$, we know $N_1+f_1\le N_1+N_2$. Finally, we can express the achievable sum-DoF as $\min\{N_1+f_1,N_2+f_2,M_1+M_2,N_1+N_2\}$ by further considering the limitation of transmitter space $M_1+M_2$.
		\item If $f_1> N_2$, which indicates that the sum-DoF is restricted to the number of receive antennas. In this sub-case, the achievable sum-DoF is $f_2+N_2+0=N_2+f_2$. Since $f_1>N_2$ and $N_1> f_2$, we know that $N_1+f_1>N_2+f_2$. Similarly, we know $N_2+f_2\le N_1+N_2$. Finally, we can express the achievable sum-DoF as $\min\{N_1+f_1,N_2+f_2,N_1+N_2,M_1+M_2\}$ by further considering the limitation of transmitter space $M_1+M_2$.
	\end{enumerate}
	
	Since $N_1+N_2 \ge \{N_2+f_2,M_1+M_2\}$ in this case, we can simplify the expression of the achievable sum-DoF as $\min\{N_1+f_1,N_2+f_2,M_1+M_2\}$. 
	
	An integer optimization problem for optimal $f_1^*$ and $f_2^*$ is formulated as%given by
	\begin{eqnarray}
		\mathcal{P}_3:	\max_{f_1,f_2 \in \mathbb{N}} && \!\!\!\!\! \!\!\!\!\! \min\{N_1+f_1,N_2+f_2,M_1+M_2\}   \nonumber \\
		\text{s.t.} && \!\!\!\!\! \!\!\!\!\! f_1N_1 + f_2N_2 \le R, f_1\in[0,M_2],f_2\in [0,M_1]. \nonumber 
	\end{eqnarray} 
 When $R\geq N_1N_2-N_2^2$, due to $\min$ function, once if $N_1+f_1=N_2+f_2$, the objective reaches the maximum. 
	Specifically, by exhausting $R$, we aim to solve the following two equations
	\begin{equation}
		\begin{cases}
			N_1+f_1 = N_2+f_2,\\
			f_1N_1+f_2N_2=R.
		\end{cases} \label{Case3-E1}
	\end{equation}
	By solving \eqref{Case3-E1} with $f_1\in[0,M_2],f_2\in [0,M_1], \, f_1,f_2\in \mathbb{N}$, the optimal $f_1^*$ and $f_2^*$ are given by
	\begin{equation}
		(f_1^*,f_2^*)=( \min \{\lfloor \gamma_5 \rfloor,M_2\}, \min\{\lfloor \gamma_6 \rfloor,M_1\}), \label{Case3-E2}
	\end{equation}
	where $\gamma_5  =   \frac{R-N_1N_2+N_2^2}{N_1+N_2} $ and $  \gamma_6  =  \frac{R-N_1N_2+N_1^2}{N_1+N_2} $. Substituting \eqref{Case3-E2} into the objective function of Problem $	\mathcal{P}_3$ gives $\min\{\lfloor \frac{R+N_1^2+N_2^2}{N_1+N_2} \rfloor,M_1+N_2,M_1+M_2\}$.
 When $R< N_1N_2-N_2^2$, however, this equivalence relationship, i.e., $N_1+f_1=N_2+f_2$, cannot hold. Due to $N_1 > N_2$, RIS is devoted to elevate bottleneck term $N_2+f_2$ in $\min$ function, leading to
	\begin{equation}
		\begin{cases}		 
			f_1N_1+f_2N_2=R, \\		
			N_1 + f_1 > N_2 + f_2 \\
			f_1 = 0, \\
		\end{cases}		
		\label{Case3-E3}
	\end{equation}
	By solving \eqref{Case3-E3} with $f_1 \in [0,M_2],\,f_2 \in [0,M_1], \, f_1,f_2\in \mathbb{N}$, the optimal $f_1^*$ and $f_2^*$ are given by
	\begin{equation}
		(f_1^*,f_2^*)=\left(0,\min \{\lfloor R/N_2\rfloor, M_1\}\right). \label{Case3-E4}
	\end{equation}
	Substituting \eqref{Case3-E4} into the objective function of Problem $	\mathcal{P}_3$ gives $\min\{N_2+\lfloor R/N_2 \rfloor,M_1+N_2,M_1+M_2\}$.
	Therefore, it can be seen that Table I Case-3 is achieved, where $R =N_1N_2-N_2^2$ satisfies $\lfloor \frac{R+N_1^2+N_2^2}{N_1+N_2} \rfloor = N_2+\lfloor R/N_2 \rfloor$.

	\subsection{Discussion}
	%		To verify the effectiveness of the proposed scheme in MIMO IC, we compare it with the MIMO IC without RIS. Additionally, $R$ is altered to investigate the impact of different numbers of RIS elements on the sum-DoF. Without loss of generality, we assume that $M_1=M_2=M$ and $N_1=N_2=N$. $R$ is fixed in Fig. \ref{DoFfig} and in Fig. \ref{DoFfig2} it holds a constant ratio to $MN$. As illustrated in Fig. \ref{DoF}, RIS can elevate the sum-DoF of the MIMO IC when $\frac{M}{N}$ is in the interval $[\frac{1}{2},2]$. Moreover, the number of RIS elements affects the sum-DoF. It is observed that more RIS elements result in a higher sum-DoF.  If $\frac{M}{N}$ is out of this interval, the maximum sum-DoF can be achieved by zero-forcing scheme and RIS is not needed.

\begin{figure}[htbp]
		\centering
		\includegraphics[width=.7\linewidth]{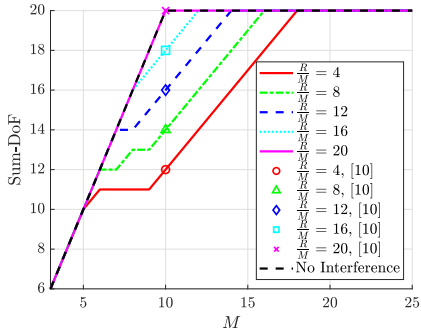}			
		\caption{$M_1=M_2=M$ and $N_1=N_2=10$.}\label{DoFfig1}
	\end{figure}

	\begin{Remark}
		Fig. \ref{DoFfig1} shows that increasing $R$ can significantly elevate the sum-DoF of two-user MIMO IC (given in \cite{jafar} or $R=0$ in Table I). This exhibits the superiority of RIS in assisting the two-user MIMO IC. Furthermore, all interference is cancelled out and the sum-DoF is $2\min\{M,N\}$ if $R\geq 2MN$, where the optimal sum-DoF is attained. Fig. \ref{DoFfig1} shows that in our settings, results of \cite{co} is a special case of ours. Additionally, our achievable sum-DoF can perfectly match the result in \cite[Theorem 4]{activepassive}, when both $\textrm{Tx}$s and $\textrm{Rx}$s are single-antenna and the number of users is $2$ in \cite{activepassive}. 
	\end{Remark}
	\begin{figure}[htbp]
		\centering
		\includegraphics[width=.7\linewidth]{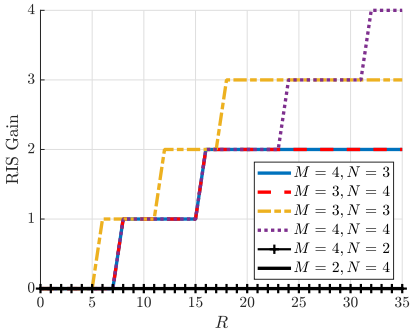}			
		\caption{RIS Gain with respect to $R$.}\label{DoFfig3}
	\end{figure}
	{\setlength{\parindent}{0cm}
		\textbf{Proposition 1} (\textit{\textbf{When RIS will help}}): \textit{ When $R \geq M_2 + M_2\mathds{1}(M_1=M_2)$, $M_2 < N_1+N_2$ for Case-1; $R\geq N_1 + N_1\mathds{1}(M_1=N_1)$ for Case-2-1, $R\geq M_1$ for Case-2-2; and $R\geq N_2+N_2\mathds{1}(N_1=N_2)$, $N_2 < M_1+M_2$ for Case-3, the derived achievable sum-DoF in Table I for an active RIS-assisted two-user MIMO IC is higher than the sum-DoF of two-user MIMO IC, where $\mathds{1}(\cdot)$ denotes the indicator function.}
		\label{Pro1}}
	
	\begin{IEEEproof}
		Please refer to Appendix \ref{pro1proof}.
	\end{IEEEproof}

	\begin{Remark}
		With an active RIS, it is shown that the existence of DoF gain depends on the relationship of $R$ and $(M_1,M_2,N_1,N_2)$. If there is no DoF gain,  rate gain will be marginal if SNR is high enough.
	\end{Remark}

	{\setlength{\parindent}{0cm}
		\textbf{Proposition 2} (\textit{\textbf{RIS gain for symmetric antenna configurations}}): \textit{
			The RIS gain is defined as the difference between achievable sum-DoF of an active RIS-assisted two-user MIMO IC and sum-DoF of two-user MIMO IC. Thus, based on Table I, for symmetric antenna configurations, i.e., $M_1=M_2=M$ and $N_1=N_2=N$, the RIS gain is given below 
			\begin{equation}
				\! 			\text{RIS Gain} = 
				\begin{cases}
					\min\{\lfloor\frac{R}{2M}\rfloor,2N-M\}, & \!\!\! N\le M<2N, \\
					\min\{\lfloor\frac{R}{2N}\rfloor,2M-N\}, & \!\!\! M< N<2M, \\
					0, & \!\!\! 2N \le M,\,\,2M \le N.
				\end{cases} \nonumber
	\end{equation}
%\begin{align}
%	 			\text{RIS Gain} &= \nonumber\\
%	&\hspace{-.8cm}\begin{cases}
%		\min\{\lfloor\frac{R}{2M}\rfloor,2N-M\}, & \!\!\! N\le M<2N\ \text{and}\ R\geq 2M, \\
%		\min\{\lfloor\frac{R}{2N}\rfloor,2M-N\}, & \!\!\! M< N<2M\  \text{and}\ R\geq 2N, \\
%		0,  &\hspace{-.475cm}\begin{cases}
%			N\le M<2N\ \text{and}\ R\leq 2M,\\
%			M< N<2M\  \text{and}\ R\leq 2N,\\
%			2N \le M,\,\,2M \le N.
%		\end{cases}
%	\end{cases} \nonumber
%\end{align}
}}
	\begin{IEEEproof}
		If $N\le M<2N$, $\min\{\lfloor \frac{R+2M^2}{2M} \rfloor,M+N,2N\} -M$ gives $\min\{\lfloor\frac{R}{2M}\rfloor,N,2N-M\}$, which is simplified to $\min\{\lfloor\frac{R}{2M}\rfloor, 2N-M\}$ due to $M\ge N$. If $M< N<2M$, $\min\{\lfloor \frac{R+2N^2}{2N} \rfloor,M+N,2M\}- N$  gives $\min\{\lfloor\frac{R}{2N}\rfloor,M,2M-N\}$, which is simplified to $\min\{\lfloor\frac{R}{2N}\rfloor,2M-N\}$ due to $M < N$.  For $2N \le M$ and $2M \le N$, it can be seen that the RIS gain is zero by $2N-2N$ and $2M-2M$, respectively. 
	\end{IEEEproof}

	\begin{Remark}
		We provide some examples of RIS gain in	Fig. \ref{DoFfig3}. It's worth mentioning that RIS gain for $2N \le M$ and $2M \le N$ is $0$ because transmit zero-forcing and receive interference decoding are enough to cancel out all interference. In addition, when the number of RIS elements is not enough to eliminate the minimum number of interference links, the RIS gain is also zero. Specifically, when $N\le M<2N$ and $R<2M$, or $M< N<2M$ and $R<2N$, the RIS gain is $0$. 
	\end{Remark}
	
%		\begin{figure}
%		\centering
%		
%		\includegraphics[width=.72\linewidth]{DoF3.eps}			
%		\caption{$M_1=M_2=M$ and $N_1=N_2=10$.}\label{DoFfig1}
%		%	\centering			
%		%	\includegraphics[width=.55\linewidth]{DoF2.eps}			
%		
%		%	\caption{$M_1= 5, M_2=3$ and $N_1=4,N_2=3$.} \label{DoFfig2}
%		%	\label{DoF}
%		
%		\centering
%		
%		\includegraphics[width=.72\linewidth]{DoF4.eps}			
%		\caption{RIS Gain with respect to $R$.}\label{DoFfig3}
%		
%	\end{figure} 
	
%	\begin{figure}
%		\centering
%		\begin{minipage}{.48\linewidth}
%			\centering
%			\includegraphics[width=\linewidth]{DoF3.eps}			
%			\caption{$M_1=M_2=M$ and $N_1=N_2=10$.}\label{DoFfig1}
%		\end{minipage}
%		\begin{minipage}{.48\linewidth}
%			\centering
%			\includegraphics[width=\linewidth]{DoF4.eps}			
%			\caption{RIS Gain with respect to $R$.}\label{DoFfig3}
%	\end{minipage}
%\end{figure}

		%	\centering			
		%	\includegraphics[width=.55\linewidth]{DoF2.eps}			
		
		%	\caption{$M_1= 5, M_2=3$ and $N_1=4,N_2=3$.} \label{DoFfig2}
		%	\label{DoF}

%	\begin{figure}
%		\centering
%		\includegraphics[width=.55\linewidth]{DoF3.eps}			
%		\caption{$M_1=M_2=M$ and $N_1=N_2=10$.}\label{DoFfig1}
%		\centering
%		\includegraphics[width=.55\linewidth]{DoF4.eps}			
%		\caption{RIS Gain with respect to $R$.}\label{DoFfig3}
%	%	\centering			
%	%	\includegraphics[width=.55\linewidth]{DoF2.eps}			
%	
%	%	\caption{$M_1= 5, M_2=3$ and $N_1=4,N_2=3$.} \label{DoFfig2}
%	%	\label{DoF}
%	
%	
%\end{figure} 
	
	\section{Conclusion}
	In conclusion, we obtained an achievable sum-DoF of an active RIS-assisted two-user MIMO IC with arbitrary antenna configurations. A new achievable scheme by integrating RIS beamforming, transmit zero-forcing, and interference decoding was given. 
	We proved that the RIS gain regrading DoF is possible when the sufficient condition is satisfied. The RIS gain was further derived for symmetric antenna configurations. 
	We find some insightful results on RIS gain. When the zero-forcing or interference decoding capability is sufficient to achieve the same sum-DoF as interference-free channels, the RIS gain is zero. Additionally, when the number of RIS elements is not enough to eliminate the minimum number of interference links, the RIS gain is also zero.
	
	\begin{appendices}

			\section{Proof of Proposition 1}\label{pro1proof}
			\textit{Case-1 ($M_1 \geq \{M_2,N_1,N_2\}$ $\text{and}$ $M_2 \ge N_1$)}: In this case, re-calling that the sum-DoF of two-user MIMO IC is  $\min\{M_2,M_1,N_1+N_2\}$ \cite[Theorem 2]{jafar}. We analyze the sufficient condition of RIS gain as follows: \begin{enumerate}
				\item If $M_2\ge N_1+N_2$, $\min\{M_2,M_1,N_1+N_2\}=N_1+N_2$, which indicates the sum-DoF is restricted to the number of receive antennas. 
				RIS cannot provide a DoF gain.
				\item If $M_2< \{M_1, N_1+N_2\}$, $\min\{M_2,M_1,N_1+N_2\}=M_2$, where for $R\ge M_2$, RIS can cancel at least one interference link, thereby RIS can provide a DoF gain; for $R<M_2$, RIS cannot provide a DoF gain. 
				\item If $M_2=M_1<N_1+N_2$, $\min\{M_2,M_1,N_1+N_2\}=M_2=M_1$, where for $R\ge 2M_2$ RIS can cancel at least two interference links, thereby RIS can provide a DoF gain; for $R<2M_2$, RIS cannot provide a DoF gain.
			\end{enumerate}

		 \textit{Case-2 ($M_1\geq N_2$ and $N_1 \ge M_2$)}:
			In this case, re-calling that the sum-DoF of two-user IC is $\min\{M_1,N_1,M_1+M_2,N_1+N_2\}$ \cite[Theorem 2]{jafar}. We analyze the sufficient condition of RIS gain as follows: 
			\begin{enumerate}
				\item If $N_1 < M_1$ (Case-2-1), $\min\{N_1,M_1,M_1+M_2,N_1+N_2\} = N_1$, where for $R \ge N_1$, RIS can cancel at lease one interference link, thereby RIS can provide a DoF gain; for $R<N_1$, RIS cannot provide a DoF gain.
				\item If $N_1 = M_1$ (Case-2-1),  $\min\{N_1,M_1,M_1+M_2,N_1+N_2\} = N_1 = M_1$,  where for $R\ge 2N_1$, RIS can cancel at least two interference links, thereby RIS can provide a DoF gain; for $R<2N_1$, RIS cannot provide a DoF gain.
				\item If $N_1>M_1$ (Case-2-2), $\min\{N_1,M_1,M_1+M_2,N_1+N_2\} = M_1$, where for $R \ge M_1$, RIS can cancel at lease one interference link, thereby RIS can provide a DoF gain; for $R<M_1$, RIS cannot provide a DoF gain.
			\end{enumerate} 
			
			\textit{ Case-3: ($N_1 \ge \{M_1,M_2,N_2\}$ $\text{and}$ $N_2 \ge M_1$)}: In this case, re-calling that the sum-DoF of two-user MIMO IC is  $\min\{N_1,N_2,M_1+M_2\}$ \cite[Theorem 2]{jafar}. We analyze the sufficient condition of RIS gain as follows: \begin{enumerate}
				\item If $N_2\ge M_1+M_2$, $\min\{N_1,N_2,M_1+M_2\}=M_1+M_2$, which indicates the sum-DoF is restricted to the number of transmit antennas. 
				RIS cannot provide a DoF gain.
				\item If $N_2< M_1+M_2$, $\min\{N_1,N_2,M_1+M_2\}=N_2$, where for $R\ge N_2$, RIS can cancel at least one interference link, thereby RIS can provide a DoF gain; for $R<N_2$, RIS cannot provide a DoF gain. 
			  \item If $N_2=N_1<M_1+M_2$, $\min\{N_1,N_2,M_1+M_2\}=N_2=N_1$, where for $R\ge 2N_2$ RIS can cancel at least two interference links, thereby RIS can provide a DoF gain; for $R<2N_2$, RIS cannot provide a DoF gain.
			\end{enumerate}

		\end{appendices}
		%\newpage
		\bibliographystyle{IEEEtran}
		\bibliography{DoFRIS}
	\end{document}